# Cryogenic System for the Cryomodule Test Stand at Fermilab

**Michael White, Benjamin Hansen, and Arkadiy Klebaner**

Fermi National Accelerator Laboratory
P.O. Box 500 MS 347
Batavia, IL 60510

E-mail: mjwhite@fnal.gov

**Abstract**. This paper describes the cryogenic system for the Cryomodule Test Stand (CMTS) at the new Cryomodule Test Facility (CMTF) located at Fermilab. CMTS is designed for production testing of the 1.3 GHz and 3.9GHz cryomodules to be used in the Linac Coherent Light Source II (LCLSII), which is an upgrade to an existing accelerator at Stanford Linear Accelerator Laboratory (SLAC). This paper will focus on the cryogenic system that extends from the helium refrigeration plant to the CMTS cave. Topics covered will include component design, installation and commissioning progress, and operational plans. The paper will conclude with a description of the heat load measurement plan.

## 1. Introduction
The Cryomodule Test Facility (CMTF) at Fermilab will be used to test cryomodules before they are placed into accelerators such as the Linac Coherent Light Source II (LCLS-II) at Stanford Linear Accelerator Laboratory (SLAC) and the Proton Improvement Plan II (PIP-II) at Fermilab. This paper will focus exclusively on the cryogenic system for the test stand dedicated to LCLS-II cryomodules. LCLS-II is an upgrade to the existing SLAC LCLS X-ray free electron laser, which will be used to observe biological molecular structures, molecular charge distributions, catalytic dynamics, and other processes at the sub-nanometer scale. By utilizing Superconducting Radio Frequency (SRF) cavities in the accelerator, the X-ray pulse repletion rate is expected to increase from 120 times per second in the LCLS to 1 million pulses per second in the LCLS-II at 4 GeV. One of Fermilab's contributions to the LCLS-II project is to fabricate and test 1.3 GHz and 3.9 GHz SRF cryomodules prior to their installation in the accelerator. At the peak of production, Fermilab is expected to test and ship one cryomodule per month to SLAC.

## 2. System Design
An overview of the plans for the entire CMTF building was presented in 2013 [1]. This section provides an overview of the cryogenic subsystems that have been commissioned or are currently in the process of being installed that will be used to operate the CMTS.

*2.1. Superfluid Cryogenic Plant*
The Superfluid Cryogenic Plant (SCP) was successfully commissioning in 2014. The SCP is designed to support multiple cryogenic test areas at CMTF, with supply circuits nominally at 2 K, 5K, and 40 K. The refrigeration capacity was measured as part of acceptance testing using a small test box housing a 2.0K helium storage vessel and heaters, which was mounted on the transfer line stub as shown in figure 1. The specified and experimentally tested refrigeration capacity of the SCP is shown in table 1.

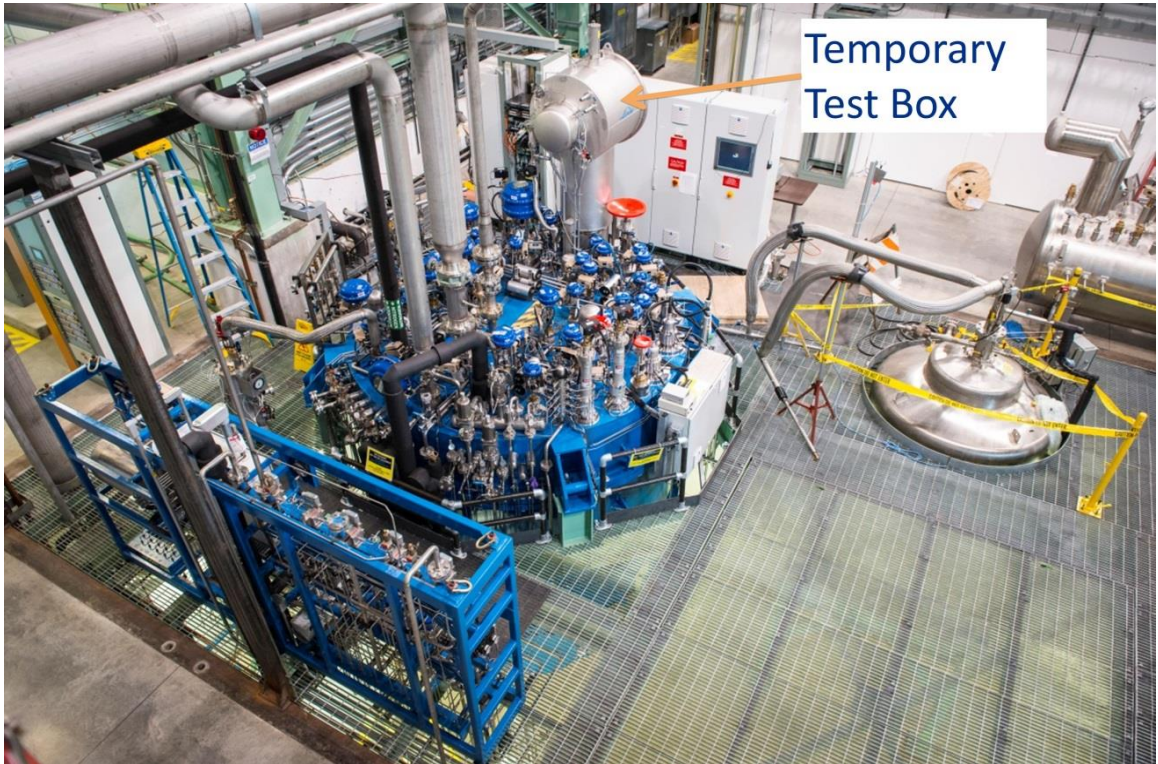

**Figure 1: Photograph of the CMTF Superfluid Cryogenic Plant along with a temporary test box. The test box was removed after SCP acceptance testing.**

**Table 1: The specified (Spec) and experimentally tested (Test) refrigeration capacity of the Superfluid Cryogenic Plant for CMTF**

| Nominal Temp | Unit | Mode 1 | | Mode 2 | | Mode 3 | |
|---|---|---|---|---|---|---|---|
| | | Spec | Test | Spec | Test | Spec | Test |
| 1.8 K | W | 250 | 257 | N/A | N/A | N/A | N/A |
| 2.0 K | W | N/A | N/A | 500 | 527 | N/A | N/A |
| 5 K to 8 K | W | 600 | 619 | 600 | 619 | 100 | 118 |
| 40 K to 80 K | W | 5,000 | 6,136 | 5,000 | 6,136 | 700 | 720 |
| Liquid He | g/s | N/A | N/A | N/A | N/A | 16 | 25 |

*2.2 Distribution Box*

The SCP test box was cut off after acceptance testing was completed. A transfer line was installed connecting the SCP to the main distribution box, which routes flow between the various test areas at CMTF. Two U-tubes are used to connect the distribution box liquid supply and vapour return to the 3000L liquid helium Dewar. Another U-tube is used to connect the cooldown return circuit of the distribution box to the SCP. A photograph of the SCP, Distribution Box, and 3000L Dewar connections is shown in figure 2. The CTI-4000, which is shown in the far right of figure 2, will also have liquid supply and vapour return U-tubes to the distribution box. However, the CTI-4000 must have an internal leak repaired before it can be commissioned.

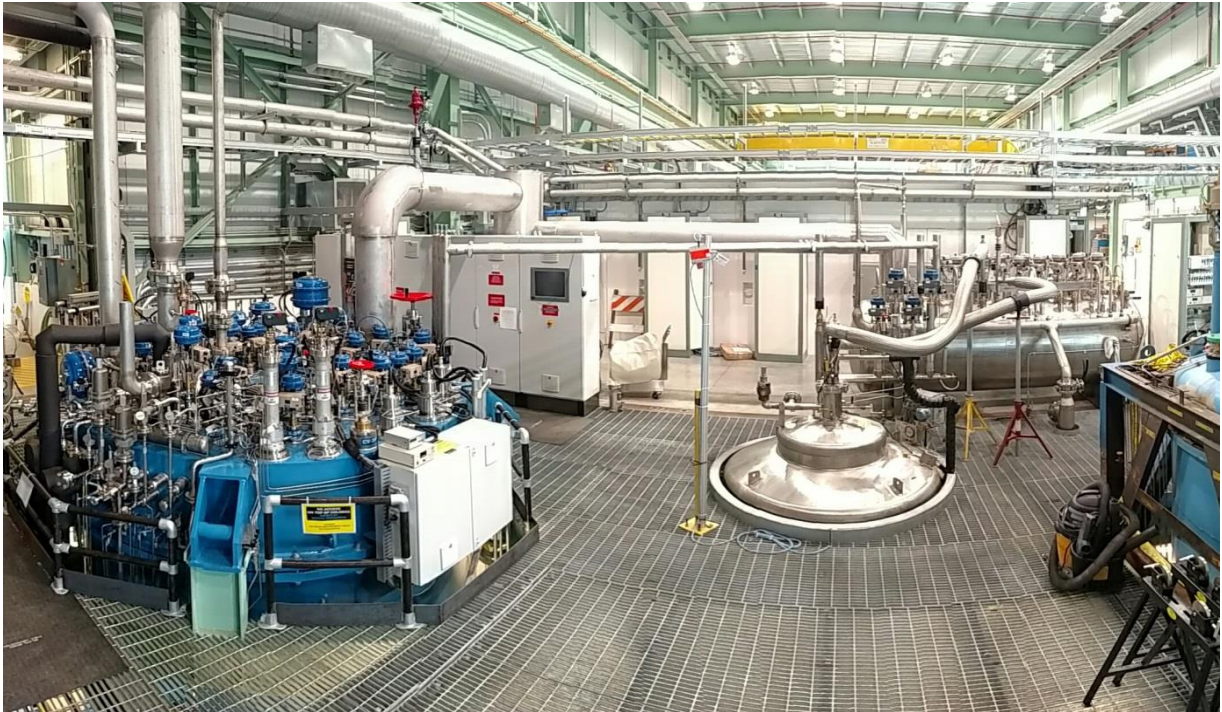

**Figure 2: Photograph showing the connections between the SCP, distribution box, and 3000L Dewar**

The distribution box has two transfer line stubs for two test areas as well as a series of supply and return bayonets for connections to a potential 3$^{rd}$ testing area. Until a third test area is built, the bayonets will serve an alternate function. Two small jumper U-tubes have been installed on the 5-8K and 40-80K circuits to allow the SCP and distribution box to remain cold regardless of the test cave temperatures. Compressor discharge pressure is automatically lowered to save on electrical power when the plant has excess capacity. Any remaining excess capacity beyond the maximum turndown of the compressor can be burned off using heaters internal to the SCP and in the 3000L dewar. The distribution box was successfully commissioned in the spring of 2015 by welding u-shaped jumpers to the internal piping of one of the transfer line stubs to simulate having a test stand connection.

*2.3 CMTS Transfer Line & Valve Box*
One of the transfer line stubs on the distribution box is dedicated to the LCLS-II Cryomodule Test Stand. Immediately after the transfer line exits the distribution box, the transfer line bends upwards and runs up into the bottom of an expansion can. This increases the transfer line elevation by 3.3 m, which allows the transfer line to pass over all the RF equipment for the two test areas at CMTF. After exiting the expansion can the transfer line runs horizontally for 21 m, then into a valve box on the roof of the CMTS cave. The opposite side of the valve box has a short horizontal run of transfer line, then drops down in the CMTS cave as shown in figure 3.

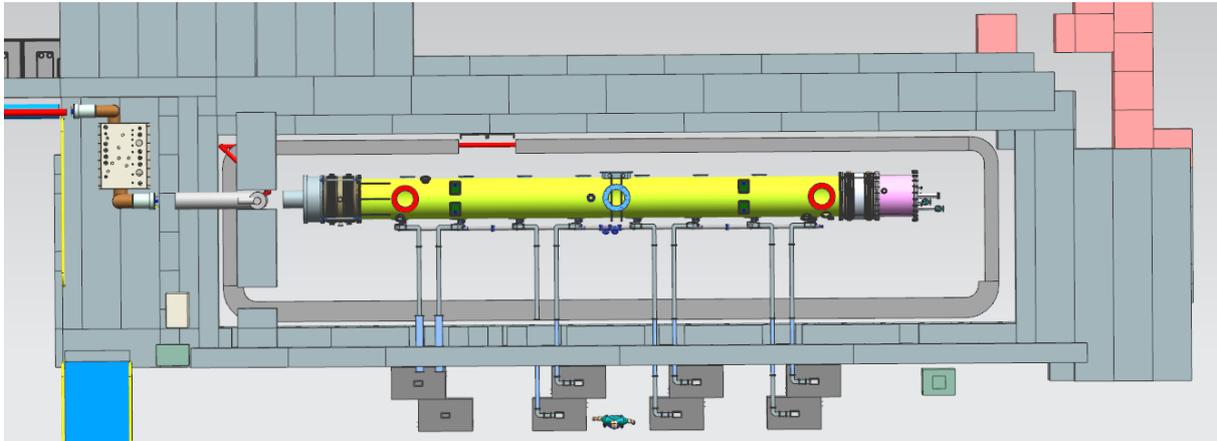

**Figure 3: Top view of the CMTS cave model with most the roof blocks removed.**

The valve box on the cave roof serves a number of functions. The most important function of the valve box is the isolation of cryogens from personnel during cryomodule installation and removal. Valves separating cryogens from personnel will be locked out and tagged for the time period between cryomodule tests. The distribution box contains most of the control valves required for cryomodule testing. However, control valves often fail to maintain a leak tight seal after extended use. A second set of manual valves are located in the CMTS valve box. The volume between the two valves can be actively pumped to ensure that no cryogens can be released into the cave. There is a vacuum break on either side of the valve box so that the vacuum space can be filled with dry nitrogen. Dry nitrogen can be injected into the valve box vacuum space to help the valves warmup and stay warm, which increases the probability the valves will seal leak tight. All pumping and backfilling of the cryomodule helium circuits will be done using ports on the valve box.

Each internal piping circuit has at least one relief valve located on the valve box. All internal piping circuits are rated for 20 bar or higher except for the cavity circuit. The niobium SRF cavities have a room temperature rating of 2.0 bar absolute and a cold rating of 4.0 bar absolute. The cryomodule cavity relief system is located on top of the section of transfer line that drops down into the cave. The cavity relief system has a cooldown valve that fails open, a relief valve sized for all room temperature failure scenarios, and a burst disk sized for all cryogenic temperature failure scenarios. The worst case cryogenic temperature relieving scenario is a sudden loss of beam tube vacuum. The air condensation heat load in a loss of beam tube vacuum was experimentally measured in a crash test at DESY [2] on similar 1.3 GHz TESLA-type cryomodule. Flow from all relief valves is collected into a low pressure helium header to minimize the chances of releasing helium into the building. In the case of the subatmospheric 2.0 K circuit, the gas collection header also helps prevent the air leaks into a cold piping circuit. The authors are unaware of any commercially available burst disks that reliably maintain helium leak tight seals over extended cycling between internal pressurization and vacuum. The plan is to have smaller high pressure burst disk inside a larger pipe with a low pressure burst disk. The space between the two burst disks is filled with helium and maintained slightly above atmospheric pressure to prevent air in-leaks.

There is a surface mounted carbon ceramic temperature sensor on every internal pipe in the valve box. In addition, two more carbon ceramic sensors are located on the thermal radiation shield for the valve box. These temperature sensors provide valuable input during warmup and cooldown of the cryomodule. The LCLS-II cryomodules have a maximum cooldown rate of 10 K/hr and a maximum inlet to outlet temperature difference of 50K on every piping circuit. Additionally, the subatmospheric return pipe has a maximum temperature difference limit of 15 K between the top and the bottom of the pipe. The subatmospheric return pipe has an inner diameter of 300 mm and is used to support the SRF cavities. Even a small amount of yielding on the 300 mm pipe can bring the cavities out of alignment, so the cooldown temperature constraints must be rigorously followed. The temperature sensors in the valve box can give operators early warning if the temperature constraints are about to be exceeded, which give the operators enough time to make corrections to stay within the temperature constraints. The valve box return side temperatures are used to verify that cryomodule is warm enough to be opened without condensing water on any MLI insulated components.

There is a single cooldown supply circuit coming from the SCP. Inside the distribution box there are valves which control flow from the cooldown supply pipe into the 40K supply, 5K supply, and 2K supply circuits. All three supply pipes have a Coriolis flow meter in the valve box, which will help operators divide flow between the three supply circuits so that all three circuits cool down evenly in a repeatable manner. The Coriolis flowmeters will be discussed in greater detail in the heat load section.

*2.4 CMTS Cave*

Four major components of the cryomodule test stand reside within the CMTS cave and are shown in figure 4: the Feedcap, the LCLS-II Cryomodule, the Endcap, and the Girder. The Feedcap and Endcap designs shown in figure 5 are similar to the design of the Feedcap and Endcap used at Fermilab Accelerator and Technology (FAST) facility at Fermilab [3,4]. The FAST Feedcap and Endcap designs are in turn are based on the Cryomodule Test Bench at DESY [5]. The detailed design and fabrication of the CMTS Feedcap and Endcap was done by Bhabha Atomic Research Centre (BARC). The Feedcap, Endcap, and Girder will be installed in the summer of 2015.

There are a few significant changes from the FAST Feedcap and Endcap. The tunnel for LCLS-II is sloped at 0.6°, so cavity circuit liquid level control cannot be managed in the same way as previous of 1.3 GHz cryomodule designs. Each cryomodule must have its own separate pipe for two phase flow to the cavities to avoid having all the liquid collect at lower elevations and all the vapour collect at higher elevations. In turn, this means that each cryomodule requires its own Joule-Thomson valve and cooldown valve. The control valves and riser pipe are placed in the middle of the cryomodule, which means that only half the length of the 300 mm subatmospheric pipe will see flow from the cryomodule JT or cooldown valves. A control valve was added to the Endcap so that flow can be circulated through the entire length of the 300 mm pipe during cooldowns and warmups.

The other major change from the FAST to the CMTS Feedcap and Endcap is the interconnects that connect cryomodule piping the Feedcap and Endcap. At FAST linear bellows were welded to form the connection. However, there is typically a misalignment between the Feedcap or Endcap piping and the cryomodule, which means the bellows pipe stubs had to be mitered. Measuring and creating the miter welds was quite time consuming. CMTS will use Conflat flanges with U-shaped assemblies with three flexhoses, as shown in figure 6. These flexhose interconnect assemblies can be installed quickly and can handle significant misalignment of the pipes while still providing for the thermal contraction of the cryomodule piping. The 300mm pipe subatmospheric return pipe does not have sufficient clearance for a Conflat flange, so an indium sealing flange will be used instead. The 300 mm pipe will still use linear bellows, but the bellows will be permanently welded to the Feedcap and Endcap. The adapter spool between the Feedcap and Endcap may have to be custom machined for each cryomodule, depending on how far the alignment is off.

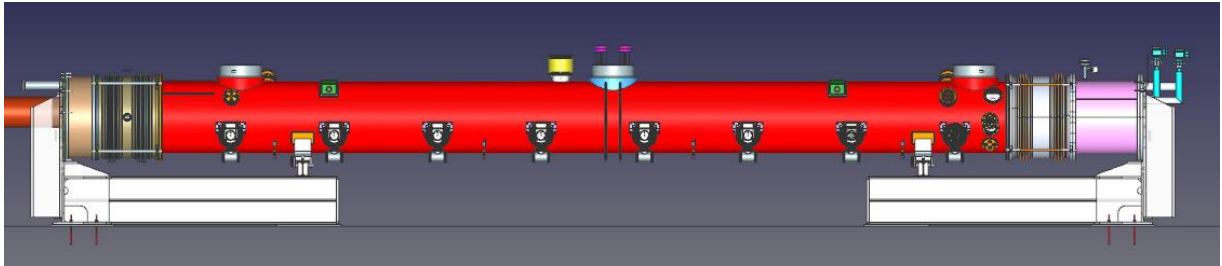

**Figure 4: Model showing a side view the Feedcap, LCLS-II 1.3 GHz cryomodule, Endcap, and Girder**

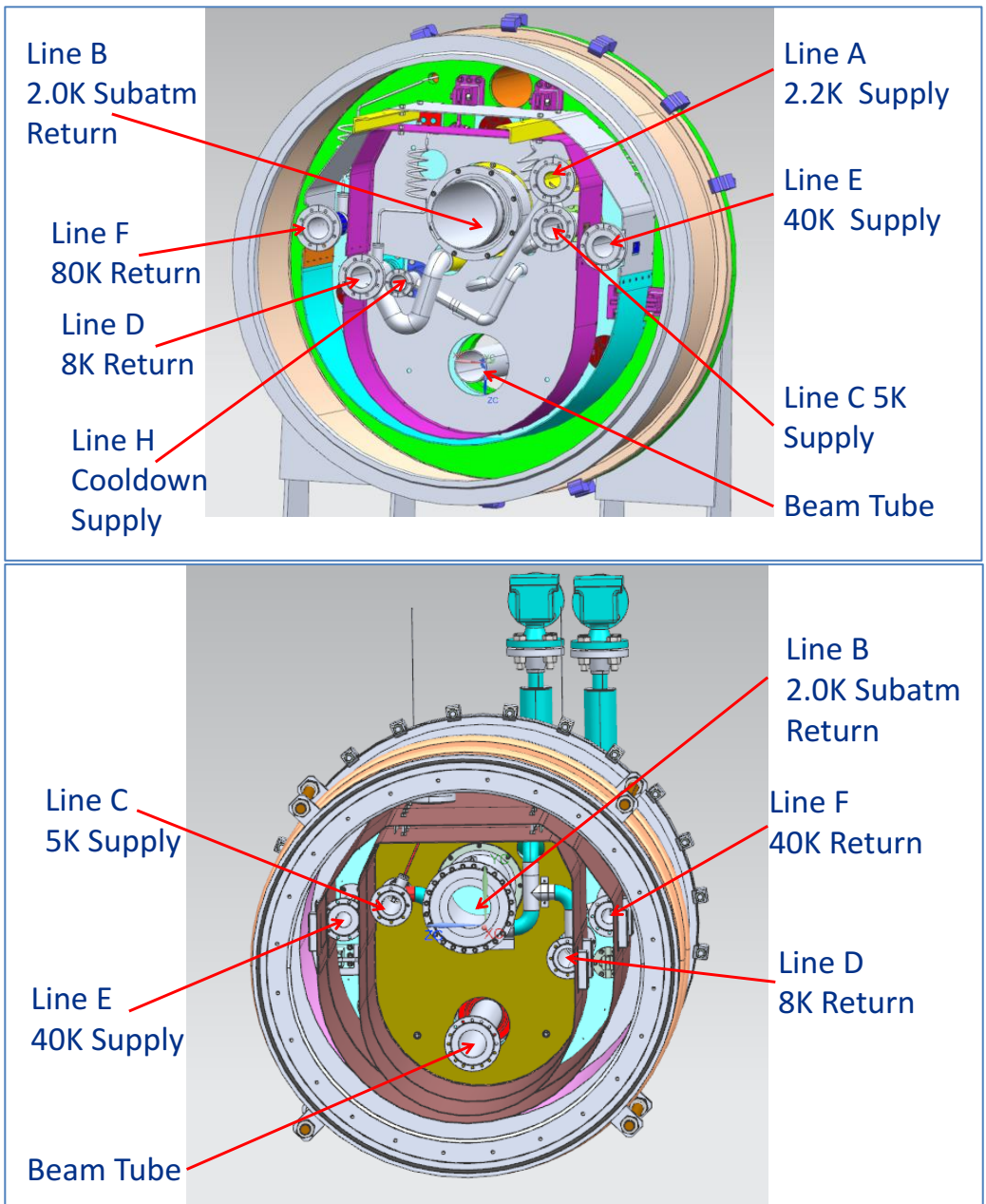

**Figure 5: Model of the Feedcap (Top) and Endcap (bottom) showing internal piping circuits**

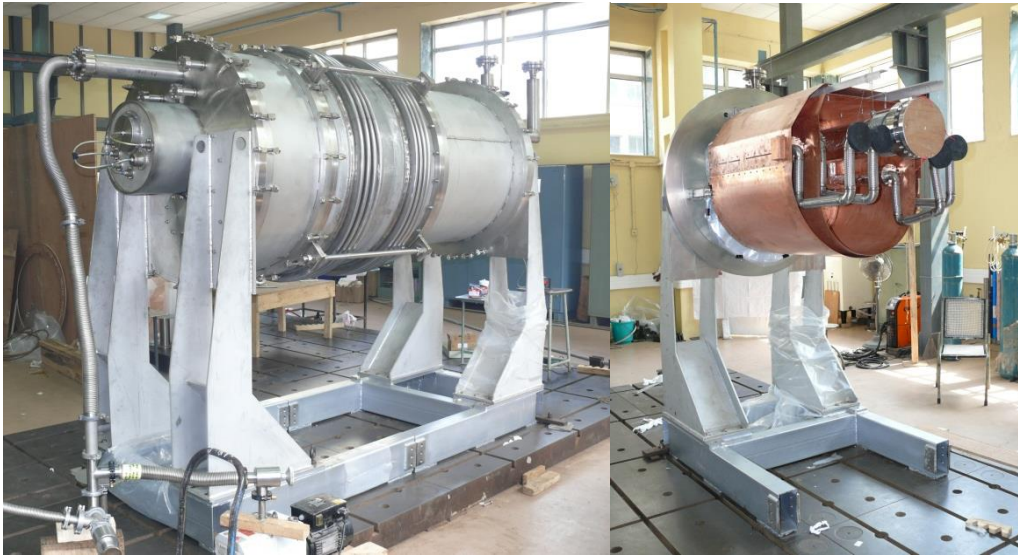

**Figure 6: Photographs of the Feedcap and Endcap at during fabrication and testing at BARC**

### 3. Heat Load Measurements

One of the more important functions of CMTS is to experimentally measure the heat loads of the LCLS-II cryomodules. The RF power will be continuous wave rather than pulsed wave for LCLS-II cryomodules, so the dynamic heat loads will be much larger than the static heat loads. The static heat loads can be estimated by scaling results from similar cryomodules used at Fermilab and other physics laboratories. The dynamic heat loads have a much greater uncertainty and are quite sensitive to the cavity quality factor $Q_0$ and the accelerating gradient $E$.

**Table 2: Summary of estimated heat loads for LCLS-II cryomodules. Static heat loads are with the RF power turned off. Dynamic heat loads are the additional heat load generated when RF power is turned on.**

| Nominal Temp | 1.3 GHz<br>$E$=16 MV/m<br>$Q_0$=2.7x10$^{10}$<br>7/8 cavities powered | | 3.9 GHz<br>$E$=12.5 MV/m<br>$Q_0$=2.0x10$^9$<br>7/8 cavities powered | |
|---|---|---|---|---|
| | Static [W] | Dynamic [W] | Static [W] | Dynamic [W] |
| 2.0 K | 7.5 | 80 | 7.5 | 100 |
| 5 K to 8 K | 22 | 15 | 22 | 15 |
| 40 K to 80 K | 110 | 130 | 110 | 140 |

To measure the flow rates through each of the circuits, a Micro Motion CMF025 mass flow meter will be used. The flow meters were originally modified for liquid helium applications by Micro Motion in collaboration with CERN for the LHC [6]. Coriolis flowmeters have several advantages over other types of flowmeters including: accuracy independent of fluid density or phase, no straight runs required upstream or downstream of the sensor, high rangeability, and low pressure drop. The flow meter transmitters have a compensation function to account for the temperature dependence of Young's modulus with stainless steel. However, the temperature measurement is performed with a platinum resistor, which will cause the transmitter to fault at liquid helium temperatures due to the low resistance of the platinum resistor. The flow transmitter is not compatible with other types of temperature sensors. There is no significant change of Young's modulus below 20K, so the Young's modulus can be set to a constant value in the flow meter transmitter without any loss of accuracy at liquid helium temperatures. The sensors are expected to have less than 1% error across the range of flow rates expected with CMTS operation. However, the error at room temperature is expected to be as high as 6.5% at room temperature due the disabled temperature compensation function. The accuracy at room temperature is much less important than the accuracy at operational temperatures. The internally mounted Cernox sensors in the Feedcap and Endcap are calibrated on-site at Fermilab and have an expected accuracy of +/- 30 mK at liquid helium temperatures. The LCLS-II heat loads will be known with a high degree of accuracy after testing at CMTS.

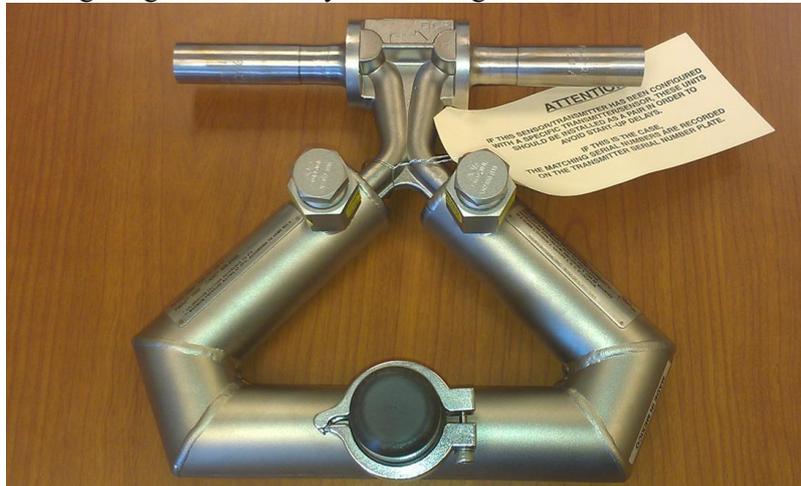

**Figure 7:** Photograph of the Micromotion CMF025 Coriolis flow meter used in the CMTS Valve Box on all three supply circuits

**Acknowledgments**
Fermilab is operated by Fermi Research Alliance, LLC under Contract No. DE-AC02-07CH11359 with the United States Department of Energy. The authors would like to thank their colleagues at Bhabha Atomic Research Centre for the design and fabrication of the CMTS Feedcap and Endcap. The authors would also like to thank the technical staff at Fermilab for all their assistance with installing and commissioning all the components described in this paper.